\documentclass{elsarticle}

\usepackage[utf8]{inputenc}
\usepackage{lineno,hyperref}
\usepackage{amssymb}
\usepackage{float}
\usepackage{epstopdf}
\modulolinenumbers[5]

\journal{Physica A}

\begin{document}
\begin{frontmatter}

\title{Is the public sector of your country a diffusion borrower? Empirical evidence from Brazil}

\author[BTS]{Leno S. Rocha\corref{cor1}}
\ead{leno.rocha@tesouro.gov.br}
\author[Utah]{Frederico S. A. Rocha}
\author[London]{Thársis T. P. Souza}
\cortext[cor1]{Corresponding author}
\address[BTS]{Brazilian Treasury Secretariat, 70.048-900 Brasilia-DF, Brazil}
\address[Utah]{Department of Computer Science, University of Utah, Salt Lake City, USA}
\address[London]{Department of Computer Science, UCL, Gower Street, London, WC1E 6BT, UK}

\begin{abstract}
We propose a diffusion process to describe the global dynamic evolution of credit operations at a national level given observed operations at a subnational level in a sovereign country.
Empirical analysis with a unique dataset from Brazilian federate constituents supports the conclusions.
Despite the heterogeneity observed in credit operations at a subnational level, the aggregated dynamics at a national level were accurately described with the proposed model.
Results may guide management of public finances, particularly debt manager authorities in charge of reaching surplus targets.
%
%
\end{abstract}


\end{frontmatter}


\section{Introduction}

Recent empirical literature supports a negative impact of public debt on economic growth in both developed and emerging markets \cite{Matesanz2015756, Panizza201421, Bua20141, doi:10.1108/JES-03-2012-0036}. 
Indeed, poorly managed public debt has been an important factor in inducing and propagating recent economic crises \cite{IMF:2014, Spilioti201534}. 

While the linkage between public debt and economic growth for national economies has long been studied, 
the investigation considering the public debt at a subnational level has been largely neglected in the literature \cite{Mitze2015208}.
Matz and Mitze (2015) \cite{Mitze2015208} showed a link between regional public debt and economic growth for German federal states. Further, 
Jenkner and Lu (2014) \cite{Lu:2014} and Buiatti, Carmeci, and Mauro (2014) \cite{RePEc:eee:jpolmo:v:36:y:2014:i:1:p:43-62} provided empirical evidence that regional
fiscal imbalance influences public debt at a national level. 

In this work, we consider a division of a sovereign state in administrative units where each one can potentially contract credit operations with the approval of the central government.
The study of aggregated regional debt dynamics is important as it impacts the public debt at the national level.
We discuss the aggregated subnational debt related to credit operations pleaded by regional governments and we propose a diffusion process \cite{shone2002economic} 
to model the number and monetary volume of such operations.
We describe mechanisms in financing that motivate the modeling of the response variables to be a sigmoid curve annually.

A case study with federated entities in Brazil confirms this sigmoid pattern and showcases the adequacy of the proposed diffusion processes. 
The applicability of the proposed methods in other sovereign countries is also considered. 
Results may be applied in distinct countries where credit pleas at the subnational level are legitimate via consent of a central public finance management, 
as is the case for the United Kingdom and Japan, for instance.

\section{Methodology}

Federated entities or administrative divisions of a sovereign state may perform credit operations for local financing purposes.
Let $N(t)$ be the aggregated number of credit operations pleaded by federated entities in a given year,
where $t \geq 0$ represents the continuous time since the first fiscal day of the year.
We propose to model the evolution of $N(t)$ as a learning curve following a diffusion process with a sigmoid trajectory.

The velocity with which entities of the public sector have credit operations plaintiffs during the year is assumed to be intrinsically related to a learning process where the number of 
credit pleas may be low at the beginning of the year further accelerating with time until it reaches a climax point. 
By the end of the year, the evolution of the number of credit operations may decelerate to respect limits and fulfill conditions established by law. 
These constraints, in addition to the credit volume limiters, lead us to propose the following diffusion process:
\begin{equation}
	\dot N(t) = g(t)\{m^n - [N(t)]^n\}/n,
\label{eqn:1}
\end{equation}
wherein the point superscript to $N(t)$ represents the time derivative of this variable; $g(t)$ is the coefficient of diffusion; $m$ represents the potential maximum number of pleas, i.e., 
the saturation level or the carrying capacity \cite{shone2002economic}; and $n$ is a parameter which characterizes the distance between $m$ and $N(t)$.


With respect to the diffusion coefficient, we assume that the number of operations pleaded 
is linearly related to the understanding of the process by the parties involved, 
so that $g(t)=wN(t)$, in which $w$ is a learning speed parameter. Hence, we solve the differential Eq. \ref{eqn:1}, as shown in Appendix A, to find the logistic function:
\begin{equation}
	N(t) = m\{1+(m-1)\exp[-wm(t-1)]\}^{-1}.
	\label{eqn:2}
\end{equation}

Considering that the course of the events concerning credit operations can last months, and the object, value, financial conditions and other characteristics of the pleas may change significantly, 
it may become more interesting to the borrower to file a new plea. 
This situation characterizes a variable mortality rate phenomenon, which was studied initially by Gompertz \cite{Gompertz}. 
Thus, we extend the logistic model from Eq. \ref{eqn:1} to incorporate the Gompertz function, as presented in Appendix B, which leads to:
\begin{equation}
	N(t) = m\exp[-b\exp(-ct)],
    \label{eqn:3}
\end{equation}
where $b = -(\ln{m})$exp$(w)$ and $c = w$. It is convenient to let the function have more degrees of freedom, using the parameters $m$, $b$ and $c$ independently, so that more adherence to the data can be achieved. The resulting model is better suited to fit sigmoid curves with different levels of acceleration and relaxation. Moreover, it presents a good compromise between descriptive power and parsimony while having only three parameters for calibration.

\section{Case Study: Credit Operations in Brazil}

Federated units in Brazil are forbidden to issue bonds \cite{Reso43}, 
but they can take loans to fund projects of their interest. In order to accomplish credit operations with financial institutions, 
federated entities must respect limits and fulfill a series of conditions determined by law. 
The Brazilian National Treasury Secretariat (STN) is in charge of checking these limits and conditions for the credit pleas. 
Contracting credit operations by federal entities in Brazil is a continuous flow process and the pleas can be sent at any time for STN's approval. 

In this context, two main general mechanisms make the time path of the Treasury’s analysis response to be a sigmoid curve annually. The first is the
need to learn the new procedures to plead credit operations and to be informed about the updated budgetary information that may change every year.
The second mechanism is related to several credit volume limiters, such as the fiscal surplus target, and the amount of money in the market available for loans.
The first mechanism shapes the process as a learning curve, in which the number of credit pleas may be low at the beginning
of the year, while the second mechanism imposes limits and therefore defines the level of saturation of the process.




Given Eq. \ref{eqn:2}, we define $N(t)$ to be the aggregated number of credit operations of Brazilian states and municipalities.
The parameter $m$ is related to credit constraints scattered in legal rules such as: 
(i) constitutional limitations for the credit operations of the public sector\footnote{Defined in the Article 52 of the Brazilian Federal Constitution of 1988 \cite{Constituicao}.}; 
(ii) limits and conditions for the credit operations of the governments \footnote{Resolution of the Senate n. 43 of 2001 \cite{Res2872}.}; 
(iii) rules of contingency for credit to the public sector \footnote{Defined in the Resolution of the National Monetary Council n. 2.827 of 2001 \cite{Res2872}.};
(iv) fiscal space remaining from the surplus target; 
and (v) discretionary limit of external operations \footnote{Defined by the External Financing Commission according to the Article 7 of the Decree n. 3.502 of the year 2000 \cite{Decreto3502}.}.

The information about the credit operations of the federate entities was collected from the Brazilian Treasury Secretariat database \cite{Sadipem}.
These data provide specific information about the date and the value of each loan thus allowing the empirical verification of the trajectory of both the number of requirements and the claimed financial volume.
The credit data is given at state and municipality granularities. 
Here, we consider the public debt of a state as the total debt of its corresponding federated unit plus the debt of its municipalities.

\begin{figure}[!h]
\centering
\includegraphics[scale=0.38]{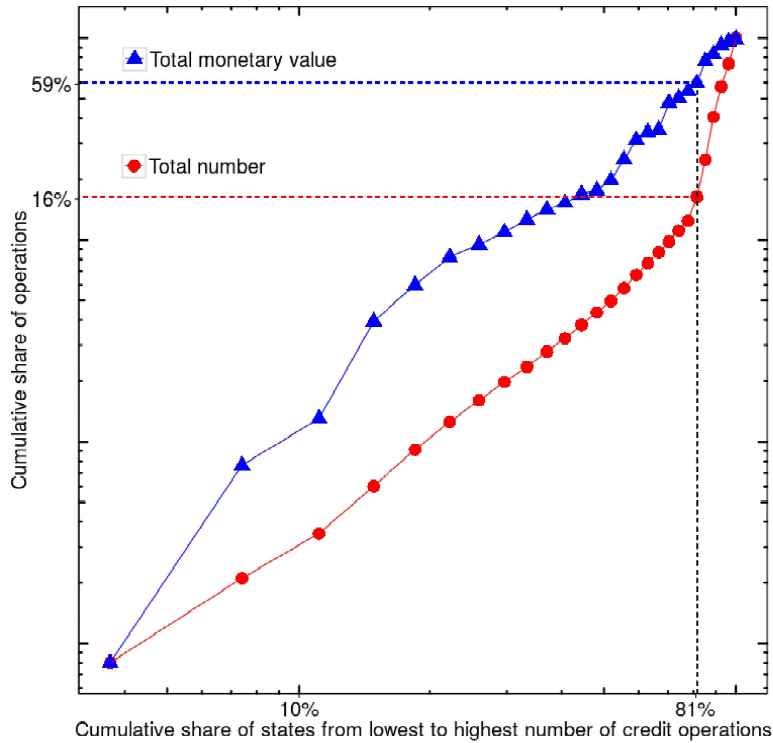}
\caption{Approximately 20\% of the Brazilian states were responsible for more than 80\% of the total number of credit operations between 2002 and 2015. The same share of states accounted for 
more than 40\% of the monetary volume in the same period.}
\label{fig:gini}
\end{figure}

\begin{figure}[!h]
\centering
\includegraphics[scale=0.25]{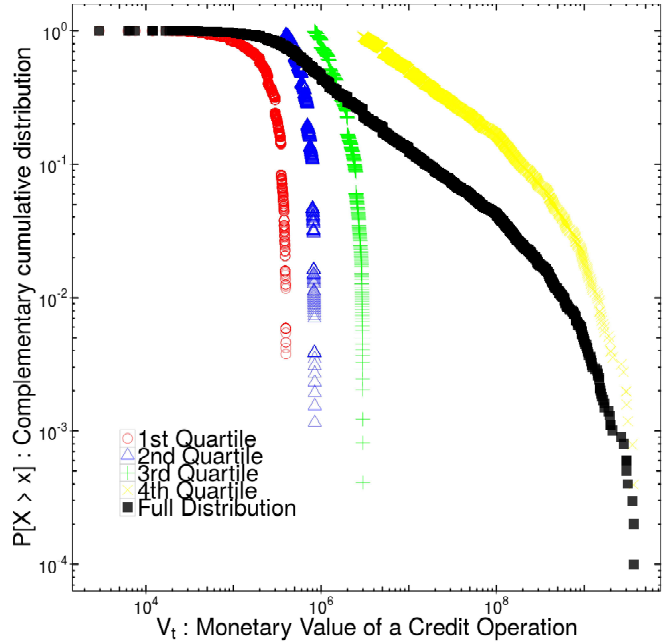}
\caption{Demonstration of heterogeneity in the value of credit operations. 
This figure shows the complementary of the cumulative distribution function of the value of credit operations in log-log scale.}
\label{fig:ecdf}
\end{figure}

Fig.~\ref{fig:gini} demonstrates that most of the operations are performed by a few states. 
A Pareto relationship is observed where 80\% of the number of credit operations came from approximately 20\% of the Brazilian states between 2002 and 2015.
An asymmetric relationship also follows in monetary terms with the same share of states being responsible for 40\% of the total pecuniary volume in the period analyzed.
The values of the credit operations also present heterogeneity. Fig.~\ref{fig:ecdf} shows the complementary cumulative distribution of the value of a credit operation performed by a state.
The distribution of values is hugely asymmetric which demonstrates the disparity among values of credit operations in the states of Brazil.

\section{Results and Discussion}

The heterogeneity observed at the state level imposes challenges in the modeling of the trajectory of the number and volume of credit operations.
Despite this complexity, we show that the observation of a sigmoid pattern at the aggregated national level results in trajectories fit with high accuracy with the proposed models.

\begin{figure}[!t]
\centering
\includegraphics[scale=0.52]{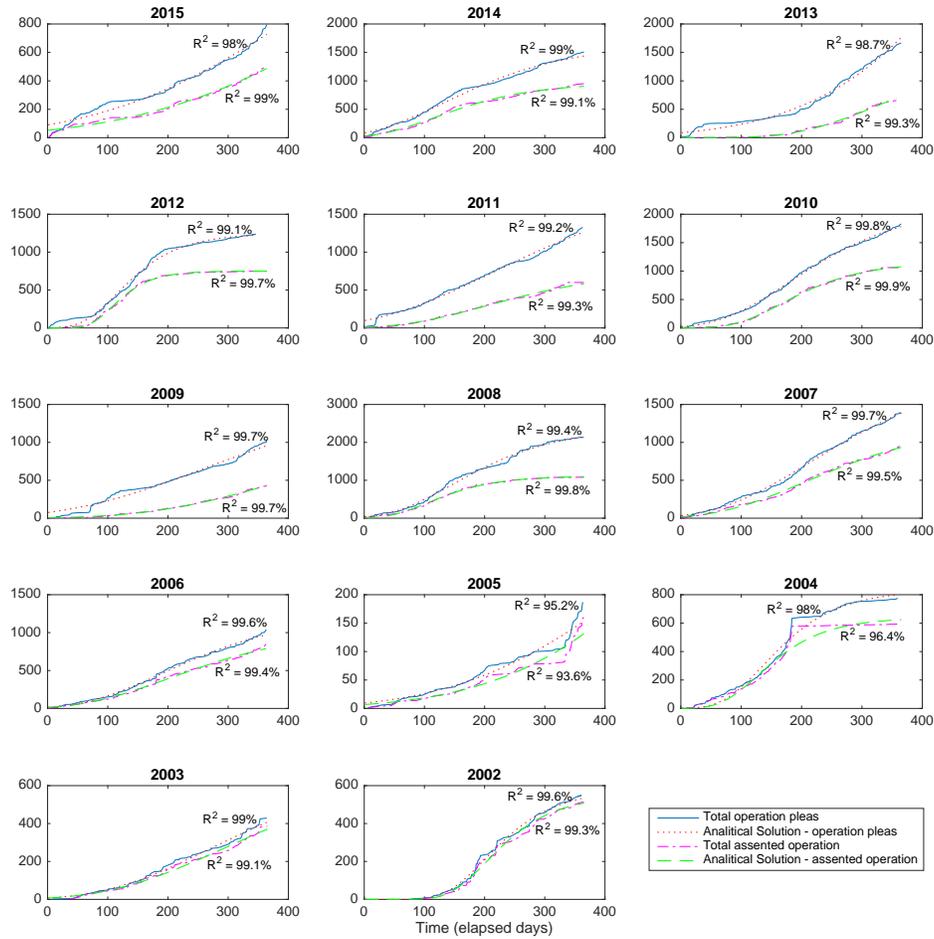}
\caption{Demonstration that the amount of public credit pleas per year follows a sigmoid diffusion process. This figure shows the total number of Brazilian credit pleas by year along with the analytic solution provided by the derived Gompertz model.}
\label{fig:diffusion1}
\end{figure}

We fit both the logistic (Eq. \ref{eqn:2}) and Gompertz (Eq. \ref{eqn:3}) models to the yearly number of credit operations.
The former model performed better for all years and its results are shown in Fig. \ref{fig:diffusion1}.
The figure shows the curves for the yearly number of credit operations that were assented by the government and the total number of credit pleas (operations rejected included).
The adherences of the logistic and Gompertz models to the data, measured by the average of the $R^2$ of the fitted curves, were above 96\% and 98\%, respectively. 
Model calibration was performed using a nonlinear least squares fitting procedure implemented in the cftool of the software Matlab \circledR. 

Notice that, in general, the yearly number of credit pleas increases slowly at the beginning of each year then it accelerates nearly reaching the yearly capacity. It then starts to slow down thus presenting a behavior well described by a sigmoid function.
In some years, the curves do not reach a flat region of saturation (e.g. 2015 and 2005), indicating that the number of pleaded credit operations might not have reached the government's budgetary expectations.


\begin{figure}[!t]
\centering
\includegraphics[scale=0.58]{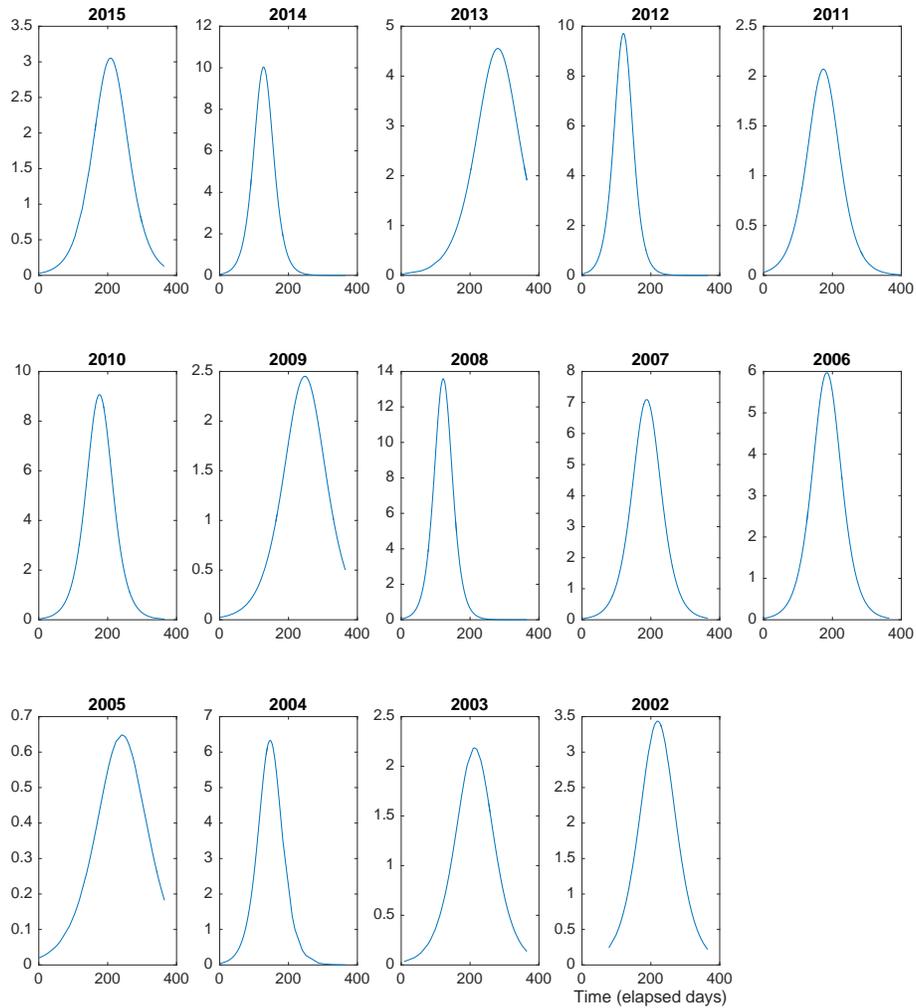}
\caption{Derivative of the number of credit operations ($\dot N(t)$). Clear peaks at the center and periods of acceleration and relaxation in the tails. Empirical derivatives estimated via central finite differences.}
\label{fig:derivative}
\end{figure}

Fig. \ref{fig:derivative} shows the derivatives of the number of credit operations ($\dot N(t)$) in time, which were fit via central finite differences.
Again, we clearly observe that each year is characterized by a period of acceleration followed by a peak that precedes an ending period of relaxation.
The peak of acceleration in the number of credit operations consistently differs by year.
Bearing in mind the small sample of these responses,
however, we observe that years with a peak at the beginning of the period tend to follow years with a peak at the end of the period and vice-versa.
On the other hand, there is no clear tendency when peaks occur in the middle of the year.
These empirical observations indicate that the aggregated number of credit operations at a subnational 
level present seasonal patterns which can be explained by political-economic cycles \cite{Sakurai2010, NBERw3830}, for instance, related to election periods.

From 2002 to 2015, the variability of credit operations pleas in monetary terms by year is considerable, especially due to outlier loans of very high value, contrasting with the wide majority of operations of small value pleaded by small states.
This dispersion prevents the usage of the model of the number of operations to describe the dynamics in monetary terms by simply multiplying the number of credit operations by the yearly mean value of the pleas.
However, we observe that the aggregated volume of credit operations in monetary terms also follows a sigmoid curve to some extent. 
This observation enables the usage of the proposed models by simply replacing the number of credit pleas $N(t)$ with its corresponding value in monetary terms.

\begin{figure}[!t]
\centering
\includegraphics[scale=0.53]{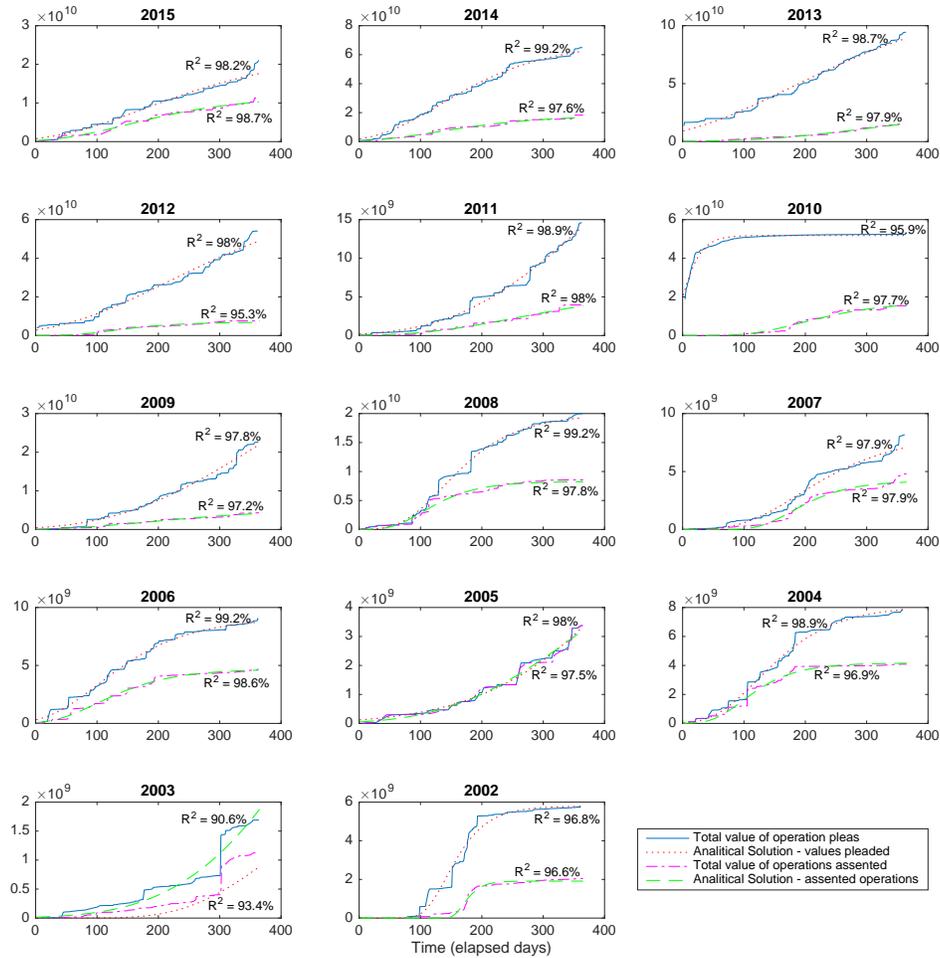}
\caption{Demonstration that the Gompertz model explains to some extent the public credit pleas also in monetary terms. Figure shows the total credit pleaded per year along with the analytical solution provided by the derived Gompertz model.}
\label{fig:diffusion2}
\end{figure}

Fig. \ref{fig:diffusion2} shows the results of the fitting further demonstrating that the sigmoid modeling can also be applied to the pecuniary aspect of the credit operations, 
though with smaller quality, since the model does not capture the ``jumps'' due to the presence of outliers in the pleaded values.

\section{Conclusions}

We proposed a diffusion process to describe the dynamic evolution of the number and volume of credit operations aggregated from a subnational level.
We uncovered a sigmoid pattern in the dynamics of credit operations further validated with empirical data from Brazilian federate constituents. 
Despite the heterogeneity observed at the subnational level, the aggregated dynamics at the national level were accurately described with the proposed diffusion processes.
Moreover, even with the particular behavior presented in each year with respect to acceleration, peak and capacity, 
all the curves were meaningfully adherent to sigmoid trajectories in terms of the number and monetary volume of credit operations.
These results enable macroeconomic policy-making without a cumbersome analysis at a subnational level.


With the calibration of the proposed diffusion model, it is possible to obtain insights into the evolution of public credit operations in a given year,
which allows improvements of economic nature, like the commitment to surplus targets. 
Possible extensions of this analysis may take into account market expectations and goals set by the government. 
Key variables to take into consideration in a Gompertz parameter prediction exercise may include: GDP,
credit risk ratings and dummy variables for elections and anti-cyclical government interventions.


The proposed model might be useful to explain credit operations of other countries.
The sigmoid pattern is likely to be observed in other sovereign states in which local government rely upon credit operations, like the United Kingdom \cite{DCLG} and Japan \cite{JFM} \cite{JLGC},
where at least the limitation on loan amounts and budgetary principle of annuity are immediate similarities that apply. 
Further research along with the availability of consolidated data on other countries would allow for the applicability and extension of proposed models.

\section*{Acknowledgments}

F.S.A.R. is grateful to CAPES (Coordination for Enhancement of Higher Education Personnel) and to the University of Utah for the financial support. T.T.P.S. acknowledges financial support from CNPq - The Brazilian National Council for Scientific and Technological Development.


\bibliography{biblio}

\appendix
\section{Solution of the logistic diffusion process}

Recall Eq. \ref{eqn:1}, the differential equation of generalized growth, and consider it with $t \in \mathbb{R}^+$ and $n \in \mathbb{N}^*$:

\begin{equation}
  \dot N(t) = \frac{dN(t)}{dt} = g(t)\{m^n - [N(t)]^n\}/n
\label{eqn:4}
\end{equation}

It is possible, using a notation abuse, to rewrite it as:

\begin{equation}
  \ \frac{ndN(t)}{N(t)\{m^n - [N(t)]^n\}} = wdt
\label{eqn:5}
\end{equation}

Making use of partial fractions, we have:

\begin{equation}
 nm^{-n} \left\{ \ \frac{1}{N(t)}+\ \frac{[N(t)]^{n-1}}{m^n - [N(t)]^n} \right\} dN(t)= wdt
\label{eqn:6}
\end{equation}

By the chain rule, $d[N(t)]^n = n[N(t)]^{n-1}dN(t)$ and performing the integration, it follows:

\begin{equation}
 n\ln|N(t)| - \ln|[N(t)]^n - m^n| = wm^n(t - t_0)
\label{eqn:7}
\end{equation}
 	
Once $m \ge N(t) \ge 0$, we can eliminate the modulus notation, and doing the exponentiation we have:

\begin{equation}
  \ \frac{[N(t)]^n}{ m^n - [N(t)]^n} = \ \frac{1}{m^n-1}\exp[wm^n(t-t_0)]
\label{eqn:8}
\end{equation}
in which $t_0$ is the initial moment, that for methodological definition is the first of the year, i.e, $t_0=1$. Finally, isolating the function of interest we have:

\begin{equation}
  N(t) = m\{1+(m^n-1) \exp[-wm^n(t-1)]\}^{\ \frac{-1}{n}}
\label{eqn:9}
\end{equation}

Assuming  $n=1$, which makes the growth expressed in the differential Equation \ref{eqn:1} to be proportional to an unidimensional distance, one arrives at the formula of Equation \ref{eqn:2}. 

\section{Derivation of Gompertz equation}

Writing powers via logarithms and making $n \to 0$, which corresponds to a growth proportional to a logarithmic distance \cite{Gompertz} in the differential Equation \ref{eqn:1} , we have:

\begin{equation}
 \lim_{n\to 0} \dot N(t) = wN(t) \lim_{n\to 0} \ \frac{\exp(n \ln m) - \exp[n \ln N(t)]}{n}
\label{eqn:10}
\end{equation}

Resorting to power series of the Napier's exponential function, it follows:

\begin{equation}
 \dot N(t) = wN(t) \lim_{n\to 0} \ \frac{1}{n} \left\{ \sum_{i=0}^{\infty} \frac{(n \ln m)^{i}}{i!} - \sum_{i=0}^{\infty} \frac{[n \ln N(t)]^{i}}{i!} \right\}
\label{eqn:11}
\end{equation}

Sorting out the two first elements of each summation, we have:

\begin{equation}
 \dot N(t) = w N(t)  \left\{ \ln \bigg[ \ \frac{m}{N(t)} \bigg] + \lim_{n\to 0}  \sum_{i=2}^{\infty} n^{i-1} \big\{ (\ln m)^i / i! - [\ln N(t)]^i / i! \big\} \right\}
\label{eqn:12}
\end{equation}

The limit of the remaining summation tends to zero, so that we have, as already stated, a logarithmic distance:

\begin{equation}
 \dot N(t) = wN(t) \ln \bigg[ \ \frac{m}{N(t)} \bigg].
\label{eqn:13}
\end{equation}

Rewriting the Equation \ref{eqn:13}, again with notation abuse, follows:

\begin{equation}
 \ \frac{d \{ \ln[ N(t) / m] \} }{ \ln [N(t)/m ]} = -w dt.
\label{eqn:14}
\end{equation}

Integrating the Equation \ref{eqn:14} we have:

\begin{equation}
 \ln | \ln[ N(t) / m] | -\ln| \ln m | = -w(t-1).
\label{eqn:15}
\end{equation}

Performing two exponentiations, we arrive at the Gompertz Equation:

\begin{equation}
  \bigg| \ \frac{\ln[ N(t) / m]} { \ln m} \bigg| = \exp[-w(t-1)]
\label{eqn:16}
\end{equation}

\begin{equation}
 N(t) = m \exp \{ -( \ln m )exp[-w(t-1)] \}
\label{eqn:17}
\end{equation}

In the original Gompertz function, $b$ = $ - $ (ln $m$) $exp(w)$ and $c$ = $w$, as can be seen in Equation \ref{eqn:17} . But as previously mentioned, we allowed the parameters to be independent in order to improve the adherence to the data. Adopting this relaxation one arrives at the 
Equation \ref{eqn:3}.

\end{document}